\documentclass[aps,pra,twocolumn,floatfix,superscriptaddress,showpacs]{revtex4}
\usepackage{graphicx}
\usepackage{dcolumn}
\usepackage{float}
\usepackage{bm}
\usepackage{amssymb,bezier,epsfig,amsmath}
\usepackage{color,epsfig}

\newcommand{\tachar}[1]{
\setbox4=\hbox{\ } \setbox3=\hbox{#1} \hbox{#1} \kern -\wd3 \kern
-\wd4 \raise 0.3\ht3 \hbox{ \vrule width \wd3 height 0.5pt} }
\input rotate
\input colordvi

\begin{document}

\title{On the role of dynamical barriers in barrierless-reactions at low energies: S($^1$D) + H$_2$}
\author{Manuel Lara\footnote{Corresponding author. E-mail: {\em manuel.lara@uam.es} \\ Present address:
Departamento de Qu\'{\i}mica F\'{\i}sica Aplicada, Facultad 
de Ciencias. Universidad Aut\'onoma de Madrid. 28049 
Madrid, Spain}} \affiliation{ Departamento de Qu\'imica 
F\'isica, Facultad de Qu\'imica, Universidad Complutense, 
28040 Madrid, Spain \\Institut de Physique de Rennes, UMR 
CNRS 6251, Universit\'e de Rennes I, F-35042 Rennes, 
France} 
\author{P. G. Jambrina}  
\affiliation{{Departamento de Qu\'imica F\'isica, Facultad 
de Qu\'imica, Universidad Complutense, 28040 Madrid, 
Spain\\Grupo de Din\'amica Molecular. Departamento de 
Qu\'imica F\'isica, Universidad de Salamanca, 37008 
Salamanca, Spain}} 
\author{A. J. C. Varandas\footnote{E-mail: {\em varandas@qtvs1.qui.uc.pt}}}
\affiliation{Departamento de Qu\'{\i}mica, Universidade de 
Coimbra, 3004-535 Coimbra, Portugal} 
\author{J.-M. Launay\footnote{E-mail:{\em jean-michel.launay@univ-rennes1.fr
}}} \affiliation{Institut de Physique de Rennes, UMR CNRS 
6251, Universit\'e de Rennes I, F-35042 Rennes, France} 
 \author{F. J. Aoiz\footnote{Corresponding author. E-mail: {\em aoiz@quim.ucm.es}}} 
\affiliation{Departamento de Qu\'imica F\'isica, Facultad de Qu\'imica, Universidad Complutense, 28040 Madrid, Spain}

\date{\today}

\begin{abstract}

Reaction probabilities as a function of total angular 
momentum (opacity functions) and the resulting reaction 
cross-sections for the collision of open shell S($^1$D) 
atoms with {\em para}- hydrogen have been calculated in the 
kinetic energy range 0.09--10 meV (1--120 K). The quantum 
mechanical (QM) hyperspherical reactive scattering method 
and quasi--classical trajectory (QCT) and statistical 
quasiclassical trajectory (SQCT) approaches were used. Two 
different {\em ab initio} potential energy surfaces (PESs) 
have been considered. The widely used RKHS PES by Ho {\em 
et al.} ({\em J. Chem. Phys} {\bf 116}, 4124, 2002) and the 
recently published DMBE/CBS PES by Song and Varandas ({\em 
J. Chem. Phys.} {\bf 130}, 134317, 2009). The calculations 
at low collision energies reveal very different dynamical 
behaviors on the two PESs. The reactivity on the RKHS PES 
is found to be considerably larger than that on the 
DMBE/CBS PES as a result of larger reaction probabilities 
at low total (here also orbital) angular momentum values 
and to opacity functions which extend to significantly 
larger total angular momentum values. The observed 
differences have their origin in two major distinct 
topography features. Although both PESs are essentially 
barrierless for equilibrium H--H distances, when the H--H 
bond is compressed the DMBE/CBS PES gives rise to a 
dynamical barrier which limits the reactivity of the 
system. This barrier is completely absent in the RHKS PES. 
In addition, the latter PES exhibits a van der Walls well 
in the entrance channel which reduces the height of the 
centrifugal barrier and is able to support resonances. As a 
result a significant larger cross section is found on this 
PES, with marked oscillations attributable  to shape 
resonances and/or to the opening of partial wave 
contributions. The comparison of the results on both PESs 
is illustrative of the wealth of the dynamics at low 
collision energy. It is also illuminating about the 
difficulties encountered in modelling an all-purpose global 
potential energy surface. 

\end{abstract}

\maketitle

\section{Introduction}

The advent of experimental techniques that allow the 
cooling of translational degrees of freedom has paved the 
way for obtaining data on reactive processes at very low 
temperatures and kinetic energies. In particular, the CRESU 
(Reaction Kinetics in Uniform Supersonic Flow) 
technique~\cite{Canosa,Lara_2010}, implemented in Rennes, 
and crossed molecular beam techniques with variable beam 
intersection-angle, as that in 
Bordeaux~\cite{Lara_2010,Lara1_2011}, can be used to 
explore reactions of atomic radicals such as F($^2P_J$), 
C($^1D_2$), O($^1D_2$), S($^1D_2$) with H$_2$ at 
translational collision energies, $E_{\rm coll}$, down to 
fractions of meV or, equivalently, few K (1K$\approx 
8.6\times10^{-5}$eV). This information can be of direct 
application to the chemistry of planetary atmospheres and 
dense interstellar clouds~\cite{personal}. Rate constants 
and excitation functions for the title 
reaction~\cite{Lara_2010,Lara1_2011} at temperatures and 
kinetic energies as low as $\sim 5$K have been already 
determined. 

The new context raises questions about the capability of 
our theoretical methodology to handle cold ($< 1$K) and 
ultracold ($< 10^{-3}$ K) processes which are sensitive to 
interactions at distances on the order of hundreds or even 
thousands of atomic units. As a simple rule of thumb in 
cold collisions, the propagation would have to be pursued 
to distances for which the potential energy is of the same 
order of the considered kinetic energy. Thereby stopping a 
calculation at distances on the order of 10 a$_0$--20 a$_0$ 
would invalidate any result for kinetic energies on the 
order of 10 K and, consequently, a major issue is the 
feasibility of a global potential energy surface (PES) to 
describe the whole configuration space in a balanced way. 
Collisions in the range $\sim 10-100$ K, as those 
considered here, may well lie in the limits of what can be 
achieved nowadays using conventional theoretical tools for 
atom--diatom system. 

Traditionally, the long-range regions of the PESs have been 
largely neglected and deemed almost irrelevant for the 
dynamics of the reaction and in the overall reactivity. 
This is so, despite the fact that a method (dubbed as 
DMBE~\cite{VAR88:255} from double many-body expansion) had 
been proposed more than two decades ago to formally 
introduce long-range forces into global potential energy 
surfaces for dynamics studies, both of the single- and 
multi-sheeted~\cite{VAR04:CI} types. Due to computational 
limitations at the time, DMBE theory assumed originally a 
semi-empirical foundation although it soon led to a general 
strategy for fitting {\em ab initio} potential energy 
surfaces encompassing both short- and long-range forces. 
Indeed, at sufficiently low kinetic energies, long--range 
interactions start playing an essential role inasmuch as 
they determine the amount of incoming flux which reaches 
the short--range region where rearrangement may occur 
\cite{Weck2006,Ospel}. While in the thermal regime 
short--range chemical forces prevail rendering the 
long--range forces less relevant, in the cold scenario 
long--range interactions become of crucial 
importance~\cite{Fara}. Interestingly, dynamical studies at 
intermediate energies in the range $\approx 1-10$ K may 
display the combined contributions of short--range and 
long--range interactions~\cite{Lara2_2011}. In this 
context,  collisions in this range of energies turn out to 
be an excellent mean to assess the quality of the PES 
sampling both long--range and short--range regions. 

The practical study of reactions at low collision energies 
is obviously restricted to barrierless reactions (or with 
barriers low enough to allow resonance enhanced tunneling). 
Among them, the slightly exoergic S($^1D$) + H$_2(X\ 
^1\Sigma_g^+)$ $\rightarrow$ SH($X\ ^2\Pi$) + H($^2S$) 
($\Delta D_e$=0.18 meV, $\Delta H_0^0=-0.29$ eV) insertion 
reaction constitutes an excellent example~\cite{aoizfe}. On 
its ground potential energy surface, the 1$^1A'$ electronic 
state, the main reaction path features a deep well 
($\approx$3.90 eV from the minimum of the reactant's 
valley. In total there are five PESs (1$^1A'$, 2$^1A'$, 
3$^1A'$, 1$^1A''$ and 2$^1A''$) that correlate with the 
S($^1$D) + H$_2$ asymptote. Of these only two, the 1$^1A'$ 
PES and the 1$^1A''$ PES, also correlate adiabatically with 
the products, whereas the remaining PESs are repulsive 
along their respective minimum energy paths connecting with 
excited states of the products. This fact does not 
precludes that other PESs could be relevant at intermediate 
regions, particularly via conical intersections that may 
call for the appearance of further non-adiabatic effects, 
especially at sufficiently high collision energies. The 
ground electronic state 1$^1A'$ has no barrier for 
perpendicular insertion whereas the first excited 1$^1A''$ 
PES has a considerably high collinear barrier 
($\approx$0.43 eV) which increases for larger angles. 
Thereby, at the moderate collision energies of the 
available experiments and moreover at low collision 
energies, the reaction is likely to be restricted to the 
ground state PES. 

The title reaction and its isotopic variants at thermal 
energies have been the subject of detailed analysis in the 
past~\cite{aoizfe,Skodje01a,CS:JPCA2001,Skodje02,Banares02,Banaresulti,HonvaLau,Rack1,Aoiztomas,Tomas,Ying,klos,Maiti,Liu98a,Liu98b,Liu00,Tian,Hankel_2009}. 
The experimental measurements by Liu and coworkers 
~\cite{Liu98a,Liu98b,Liu00} motivated a number of 
theoretical studies. In particular, Zyubin {\em et al} 
\cite{Skodje01a} carried out extensive MRCI {\em ab initio} 
calculations with multiconfiguration self--consistent field 
(MCSCF) reference wave functions for all the PESs that 
correlate with the reagents. Subsequently, an improved 
version of the 1$^1A'$ PES based on the same {\em ab 
initio} points was produced by Ho {\em et al.} using the 
reproducing kernel Hilbert space (RKHS) interpolation 
method. The resulting PES, henceforth called RKHS PES, has 
no barrier for insertion but exhibits a late collinear 
barrier whose height is $\approx$0.36 eV. As will be seen, 
this barrier plays a non-negligible role in the dynamics at 
low $E_{\rm coll}$. This PES has been extensively used in 
quasi--classical trajectory (QCT) 
\cite{CS:JPCA2001,Skodje02,Banares02,Banaresulti}, time 
independent quantum mechanical (TI-QM) 
\cite{HonvaLau,Banares02}, wave packet 
\cite{Ying,Tian,Hankel_2009}, rigorous statistical quantum 
mechanical (SQM) \cite{Rack0,Rack1,Tomas} and statistical 
quasi--classical trajectory (SQCT) \cite{Aoiztomas} 
dynamical calculations. 

In previous studies \cite{Lara_2010,Lara1_2011,Lara2_2011}, 
accurate, fully converged QM dynamical calculations were 
performed to explore the reactive behavior of the title 
reaction at low kinetic energies and to compare their 
results with the recent measurements by Sims and coworkers 
and Costes and coworkers \cite{Lara_2010,Lara1_2011}. These 
QM calculations were carried out on the RKHS ground $1^1A'$ 
adiabatic PES. However, since the accent was put on the 
accurate reproduction of the experimental data, it was 
found that the short--range description of the RKHS PES 
required to be complemented with an {\em ad hoc} 
modification of the long--range interactions. Although the 
RKHS had been widely used in the past and its short-range 
region was tested by comparison with experiments at higher 
energies \cite{Banares02,Banaresulti}, its long--range 
potential was not accurate enough to describe the 
experimental results at kinetic energies $\lesssim 10$ K, 
even though its effect was found negligible at higher 
energies. It should be recalled that the long-range 
behavior in the system is characterized by the presence of 
a significant quadrupole-quadrupole contribution that 
varies as $R^{-5}$ ($R$ being the atom--diatom distance) 
which is due to the open shell nature of the excited 
electronic state of the S atom. This potential term may 
lead to important reorientation effects at low collision 
energies. 

Recently, a new {\em ab initio} PES for the system, named 
DMBE/CBS was calculated by Song and Varandas~\cite{Var} by 
fitting accurate multireference configuration interaction 
energies with large basis sets (Dunning's aug-cc-pVTZ and 
aug-cc-pVQZ) and extrapolation to the complete basis set 
(CBS) limit \cite{VAR07:244105}. Special care was paid in 
the fitting procedure to the long range behavior, although 
the quadrupole-quadrupole electrostatic interaction was not 
explicitly considered. Given the high level of {\em ab 
initio} calculations used in its construction, one can 
expect a higher degree of accuracy as compared with 
previous PESs for the ground $1^1A'$ state of the SH$_2$ 
system. The few dynamical studies available on this PES at 
the time \cite{Var} pointed at different dynamical effects 
to those calculated on the RKHS PES \cite{Skodje02} and 
even with those reported for a similar PES by the same 
authors \cite{Var,Var2} to be addressed further below. In 
particular, the values of their QCT thermal rate constants 
at 300 K for different isotopes and intramolecular and 
intermolecular kinetic isotope effects were presented in 
ref.~\cite{Var}, indicating some discrepancies in the 
results from the previous PESs. Although, shortly after its 
publication, it was recognized  by the authors that the 
tabulated experimental branching ratios (seventh column of 
Table~4 in Ref.~\cite{Var} should read as the inverse 
values of the ones actually given), the above finding 
prompts the question about their quality and topographical 
differences. As commented on above, the analysis of low 
collision energy results is probably the most stringent 
test to assess the overall performance of a PES. Under such 
conditions the parallel study of the dynamics of a given 
reaction on two different PESs may serve to disentangle 
more or less subtle effects due to specific features of the 
potential. In the present study, only the original RKHS PES 
and DMBE/CBS PES, as they were released without 
modifications, will be employed. In any case, the 
modification of the long--range part of the RKHS PES used 
in ref.~\cite{Lara_2010,Lara2_2011} only affects the 
excitation functions at energies below 1 meV as it was 
preeviously shown \cite{Lara2_2011}. 

Shortly after the publication of the DMBE/CBS PES, another 
PES, based essentially on the same set of {\em ab initio} 
data, was released by Varandas and coworkers: the DMBE/SEC 
PES \cite{Var2}, so called for the use of the scaled 
external correlation aimed to extrapolate to the complete 
basis set and full configuration-interaction 
limits~\cite{VAR89:4379}. A comparative study of the 
dynamics on both the DMBE/CBS PES and DMBE/SEC PES has been 
recently carried out \cite{MAR011} using accurate 
time-dependent wave packet (TD--WP) calculations. That  
work was performed independently and in parallel to the 
present one that uses different theoretical methodologies, 
SQCT, QCT and rigorous TI--QM, the latter more accurate 
than TD-WP for low collision energies. The present work is 
focussed on detailed comparison of the DMBE/CBS and RKHS 
PESs exclusively at low collision energies and to 
disentangle the various dynamical effects.

At lower energies in the reactant valley lie the adiabatic 
triplet PES that correlates with the S($^3 P$)+H$_2$ and 
crosses the $^1A'$ PES in the H$_2$S well, leading to the 
same asymptote in the product's valley. Theoretical 
calculations\cite{Maiti,Tian} conclude that although the 
contribution to the product's formation via inter system 
crossing is small, the electronic non--reactive quenching 
process, S($^1D$) + H$_2$ $\to$ S($^3P_{0,1,2}$) + H$_2$, 
may play a significant role in the absolute removal of 
S($^1D$), at least at energies as low as $250$K (21.5 meV). 
The real contribution of the quenching to the collision 
process at low energies has not been cleared up by the 
recent experiments on the system and the resulting 
theoretical analysis~\cite{Lara_2010,Lara1_2011}. The 
branching ratio reaction to quenching in the considered 
range remains unknown. Hereinafter, we will consider purely 
adiabatic collisions on the ground state PES since the 
purpose of the present work is to isolate the effects of 
the features of the ground state potential on the low 
energy collision dynamics. The excellent agreement between 
recent experimental data and adiabatic calculations in the 
low energy range \cite{Lara_2010,Lara1_2011} indicates that 
the possible non--adiabatic effects mentioned above should 
be largely irrelevant for the measured dynamical 
observables (cross sections and rate coefficients) at 
sufficiently low collision energies. 

The paper is structured as follows. In the next section, we
will briefly describe the three dynamical methodologies used.
The results of the calculations will be shown in Section III
and will be discussed in Section IV. Finally, a summary of the
work and some conclusions will be given in Section V.

\section{Dynamical Methodology }

\subsection{Quasi-Classical trajectory method}

QCT calculations have been carried out for the 
S($^1D$)+H$_2$($v$=0,$j$=0) reaction by running $10^6$ 
trajectories in the 0.5 meV--10 meV collision energy range 
on each of the PES considered in this work. The integration 
step was chosen to be $4 \times 10^{-17}$ s on the RKHS PES 
and  $6 \times 10^{-17}$ s on the DMBE/CBS PES. This 
guarantees a total energy conservation better than 1 part 
in 10$^4$. Due to the small collision energies and the long 
range interaction in the potential, the initial and final 
atom--diatom distance was chosen to be 30 \AA. The detailed 
QCT procedure has been discussed elsewhere 
\cite{AHS:JCP92,ABH:JCSFAR98} and will not be repeated 
here. 

The same methodology was employed to calculate the 
excitation functions for the S + H$_2$($v$=1,$j$=0) 
reaction by running $10^6$ trajectories in the collision 
energy range between 5 meV and 500 meV. The initial 
atom-diatom distance was set to 10 \AA. To improve the 
accuracy at low values of $E_{\rm coll}$, an extra batch of 
10$^6$ trajectories was run in the energy range 0.5 meV--80 
meV on the RKHS PES. Additional batches of $10^5$ 
trajectories were also run at discrete collision energies 
(5 meV, 6.29 meV, 10 meV and 30 meV for the reaction with 
H$_2$($v$=0,$j$=0) and 32 meV and 82 meV for the reaction 
with H$_2$($v$=1,$j$=0)) using the standard QCT methodology 
\cite{AHS:JCP92} to determine the opacity functions. 

The effect of the different types of binning for the 
assignment of final states on the cross section and opacity 
was also studied. Since no resolution in final states is  
carried out in this work, the results obtained were 
insensitive to the particular binning method that could be 
used, as also found elsewhere~\cite{Var2} using a related 
approach

\subsection{Statistical Quasi-Classical trajectory method}

The SQCT method  has been described in previous 
publications \cite{ASGM:JCP07,AGS:JCP07,Aoiztomas}. It is 
equivalent in all aspects to its QM version, SQM 
\cite{Rack1}, with the sole difference that trajectories 
instead of wave functions are propagated. In the SQCT 
method, the trajectories are integrated until they reach 
the well, characterized for a negative limiting value of 
the potential (measured from the bottom of the reactant's 
valley). Calculations for the title reaction had been 
already performed using the SQCT and SQM approaches 
\cite{Aoiztomas} yielding almost identical results. In this 
case, the limiting value was chosen to be -0.6 eV for the 
reagents channel and -0.8 eV for the products. 

Calculations were performed at several values of $E_{\rm 
coll}$ over the 1 meV--10 meV and 30--280 meV ranges for 
the S($^1D$)+H$_2$($v$=0,$j$=0) and 
S($^1D$)+H$_2$($v$=1,$j$=0) reactions, respectively. 
Batches of 10$^5$ trajectories were run for each energy and 
chemical rearrangement on each of the PESs. The integration 
step size was chosen to be $7\times 10^{-17}$ s (for both 
PES) enough to ensure a conservation in the total energy 
better than 1 part in 10$^5$.

\subsection{The Quantum mechanical hypersperical approach}
\label{QMmethod}

The hyperspherical quantum reactive scattering method 
developed by Launay {\em et al.}~\cite{Launayfirst} was 
described in previous works in the context of thermal 
reactive scattering \cite{hon04}. Recent modifications of 
the method performed in order to allow the accurate 
inclusion of small anisotropic long-range interactions in 
the PES were described in depth in ref.~\cite{Lara2_2011}. 
In what follows, we will simply recall the basic concepts 
referring to previous works for more details. 

In the hyperspherical quantum reactive scattering method, 
developed by Launay~\cite{Launayfirst}, the configuration 
space is divided into inner and outer regions. The 
positions of the nuclei in the inner region are described 
in terms of hyperspherical democratic coordinates. The 
logarithmic derivative of the wavefunction is propagated 
outwards on a single adiabatic PES. At a large enough value 
of the hyper-radius the former is matched to a set of 
suitable functions, called asymptotic functions, to yield 
the scattering S-matrix. The asymptotic functions provide 
the collision boundary conditions. In fact, when working at 
thermal energies, they are the familiar regular and 
irregular radial Bessel functions which account for the 
presence of the centrifugal potential at large 
intermolecular separations. They were recently generalized 
in order to include also the effect of anisotropic 
long-range interactions which act on the reagents while 
they approach each other. This enables the study of cold 
and ultracold collisions, very sensitive to the long-range 
part of the PES, without the need of extending the inner 
calculation, in hyperspherical coordinates, to very large 
values of the hyper-radius~\cite{Lara2_2011}. 

In the current study, we choose an adiabatic treatment of
the dynamics, assuming that the collision occurs only on
the ground adiabatic PES  which will be labeled by
$V^0(R,r,\theta)$. Using the set of Jacobi coordinates
$(R,r,\theta)$ corresponding to the S+H$_2$ arrangement,
the nuclear Hamiltonian can be expressed as
\begin{eqnarray}
{\hat H} =- \frac{\hbar^2}{2 \mu} \frac{1}{R} \frac{\partial^2}{\partial
R^2} R  +\frac{1}{2 \mu
R^2} {\bm l}^2- \frac{\hbar^2}{2 m} \frac{1}{r} \frac{\partial^2}{\partial
r^2} r  +\frac{1}{2 m
r^2} {\bm j}^2.
\label{hamil}
\end{eqnarray}
Let us label with ${\bm l}$ the orbital angular momentum of 
the atom with respect to the center of mass of the diatom, 
and with ${\bm j}$ the rotational angular momentum of the 
latter. The total angular momentum of the nuclei (conserved 
in an adiabatic approach) is given by 
$\bm{J}=\bm{j}+\bm{l}$. A convenient basis in order to 
expand the nuclear wavefunction in the long-range region is 
that characterized by quantum numbers $(J,M,v,j,l)$, with 
($v,j$) the rovibrational quantum numbers of the diatom, 
$l$ the relative orbital angular momentum and ($J,M$) the 
total angular momentum  and its projection on the 
Space-Fixed (SF) $Z$ axis (represented as $\varphi^{J M}_{v 
j l }$). Such a SF basis set is used in the hyperspherical 
approach to expand the asymptotic wavefunctions, which are 
matched with the short--range information obtained in 
hyperspherical coordinates. 

If the system approaches collision with quantum numbers
$(J,M,v_0,j_0,l_0)$, we will assume  that (in addition to
$J$ and $M$) the rovibrational quantum numbers,
($v_0,j_0$), remain well conserved in the long-range region. This
is justified given the large energy gap between different
rovibrational states relative to the small considered
collision energies. Within this approximation the nuclear
wavefuncion, $\Psi^{J M}_{v_0 j_0 l_0 }$, can be expanded
in the long-range region as
\begin{eqnarray}\label{eq:desar2}
\Psi^{J M}_{v_0 j_0 l_0 }=\sum_{l} \frac{F^{l_0}_l(R)}{R} \varphi^{J M}_{v_0 j_0  l},
\label{expa}
\end{eqnarray}
where all the ``conserved" quantum numbers have been 
suppressed in the notation of the radial coefficients, 
$F^{l_0}_l(R)$. Introducing the expansion (\ref{expa}) into 
the time-independent Schr{\"o}dinger equation associated 
with a total energy $E$, $H \Psi=E \Psi$, and using the 
Hamiltonian in Eq.~\ref{hamil}, it is straightforward to 
obtain the following system of coupled radial equations 
\begin{widetext}
\begin{eqnarray}
\left[ - \frac{\hbar^2}{2 \mu}
\frac{\partial^2}{\partial
R^2}  + \frac{l(l+1) \hbar^2 }{2 \mu R^2} -E_{\rm coll} \right] F^{l_0}_{l}(R)=
- \sum_{ l' } \langle  V \rangle_{ l, l'} (R) F^{l_0}_{ l'}(R)
\label{eq:acoplamien}
\end{eqnarray}
\end{widetext}
where the collision energy, $E_{\rm coll}=E-E_{v,j}$, where 
$E_{v,j}$ is the internal energy of the diatom. $\langle V 
\rangle _{ l, l'}(R)$ designates the matrix elements in the 
SF basis of  $V^0(R,r, \theta)-V_{\rm H_2}(r)$, with 
$V_{\rm H_2}(r)$ the asymptotic H$_2$ diatomic potential. 
By inwards integration of Eq.~\eqref{eq:acoplamien} we 
obtain the ``regular'' ($F^{(1) l_0}_{l}$) and 
``irregular'' ($F^{(2)l_0}_{l}$) asymptotic radial 
wavefunctions, corresponding to an incoming 
$(J,M,v_0,j_0,l_0)$ channel~\cite{Lara2_2011}. Let us note 
that, as can be seen in Eq.~\ref{eq:acoplamien}, $\langle  
V \rangle_{l_0,l_0}(R)+ l_0(l_0+1) \hbar^2 /2 \mu R^2 $ 
(diagonal part) has the meaning of the effective potential 
felt by the colliding partners at a distance $R$ when 
approaching in the state $\varphi^{J M}_{v_0 j_0 l_0}$. 
Such meaning will be very useful below. 

Regarding the calculation of the potential matrix $\langle 
V \rangle_{ l, l'} (R)$, it is convenient to calculate the 
potential matrix associated to a basis labeled by the 
projection $\Omega_j$ of ${\bm J}$ on the Body--Fixed (BF) 
coordinate system whose $z$-axis is chosen along the Jacobi 
$\bm R$ vector. This BF basis set is given by 
\begin{eqnarray}\label{eq:jotas}
\phi^{J M}_{v  j  \Omega_j }=\frac{\chi_{v,j}(r)}{r}
\sqrt{\frac{2J+1}{4\pi}}D_{M \Omega_j}^{J*}(\alpha,\beta,\gamma)Y_{j
  \Omega_j}(\theta,0),
\end{eqnarray}
where $\chi_{v,j}(r)$ is the radial rovibrational wave 
function, $Y_{j \Omega_j}(\theta,\phi)$ the spherical 
harmonics, and the $D_{M \Omega_j}^{J*}$ denotes a Wigner 
rotation matrix element where $(\alpha,\beta,\gamma)$ are 
the Euler angles corresponding to the transformation 
between SF and BF frames. It is easy to see that the matrix 
elements of the ground PES in this basis  are thus given by 
\begin{widetext}
\begin{equation}\label{clarito}
\langle  V \rangle_{\Omega_j, \Omega_j'}(R)= \delta_{\Omega_j  \Omega_j'} 2 \pi  \int  \chi_{v,j}^2(r)
 Y_{j \Omega_j}^2(\theta,0) V^0(R,r,\theta) \sin\theta\; dr\; d\theta
\end{equation}
Once the potential matrix elements are calculated in BF 
frame, we change to the S-F basis, which basically involves 
a combination with $3j$ symbols, thus obtaining 
\begin{eqnarray}\label{camio}
\langle  V \rangle_{l,l'}(R) =  (-1)^{l+l'} \sqrt{2l+1}\sqrt{2l'+1} 
 \sum_{\Omega_j} \left(\begin{array}{ccc} j & l & J  \\ \Omega_j & 0  & -\Omega_j \end{array}\right)
\left(\begin{array}{ccc} j & l' & J  \\ \Omega_j & 0  & -\Omega_j \end{array}\right)  \langle  V \rangle_{\Omega_j, \Omega_j}(R). \nonumber
\end{eqnarray}
\noindent
\end{widetext}

Finally, let us note that we have chosen an intermolecular 
separation of $\sim 10$ a$_0$ for the matching of the inner 
information with the asymptotic functions. The S-matrix has 
been calculated on a grid of 120 equally spaced energies in 
the collision energy range 0.09$-$10 meV (that is, for 
every 'integer' energy in the range from 1 to 120 K). 
Eighteen partial waves ($J=0-17$) were required in order to 
converge the cross-sections in that energy range. Other 
convergence parameters used in the current work are the 
same which were used in the study of the title collision at 
thermal energies, and are given in 
ref.~\cite{Banares02,Banaresulti}. 

\section{Results}

Reaction cross-sections as a function of the collision 
energy (excitation function), $\sigma_R(E_{ \rm coll})$, 
for the S($^1$D) + H$_2$ ($v$=0,$j$=0) collisions in the 
center--of--mass kinetic energy range 0.09--10 meV (1--120 
K), have been calculated using the three methodologies 
described in the previous section. The calculations have 
been carried out on both the RKHS PES \cite{Skodje02} and 
DMBE/CBS PES \cite{Var} versions of H$_2$S $^1$A$'$ ground 
state PES. As stated above, the effects of singlet excited 
PESs and spin--orbit coupling are not considered here 
\cite{Lara2_2011}.

\begin{figure}[h]
\begin{center}
\includegraphics[width=1.0\columnwidth]{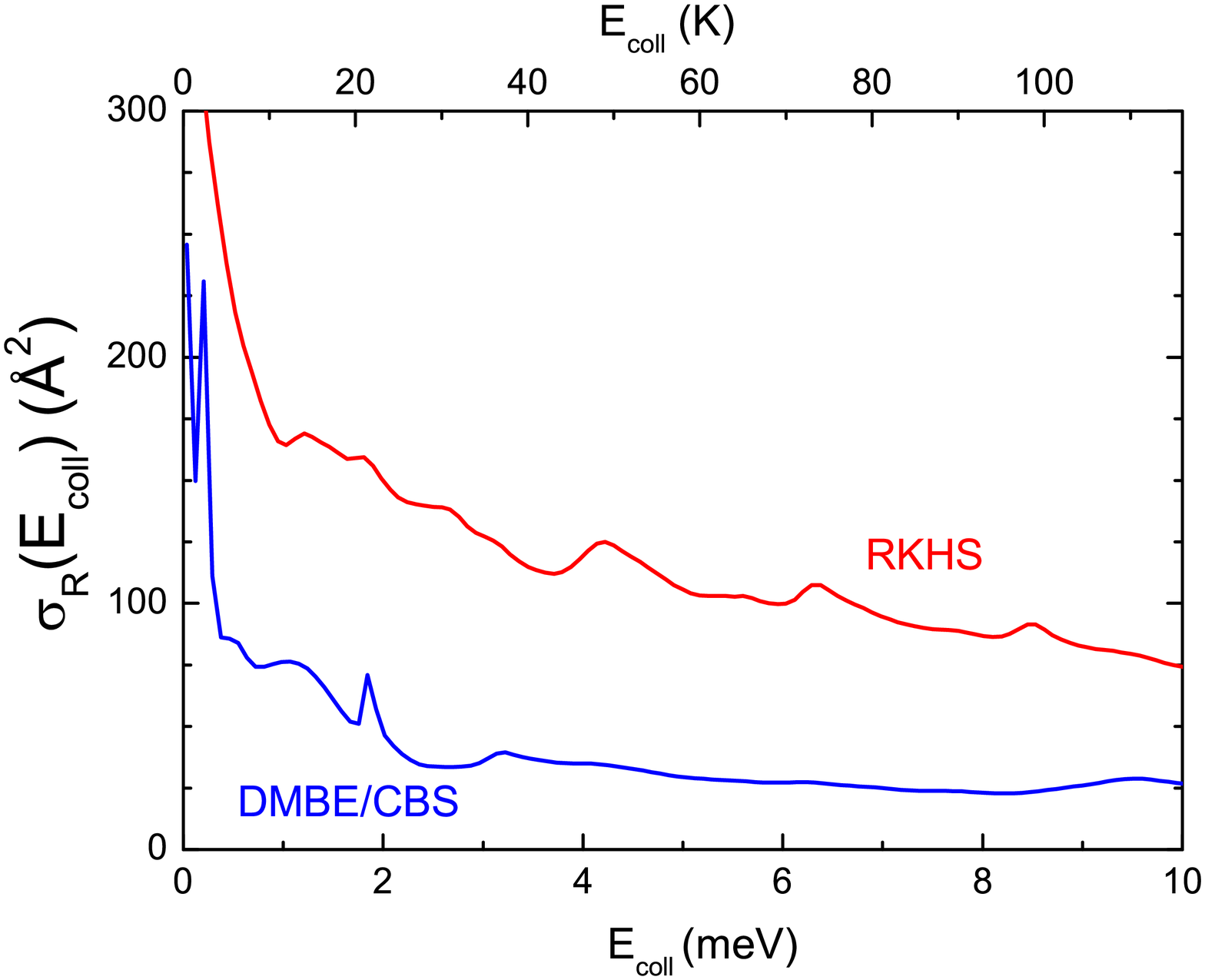}
\caption{(color online) Comparison of the QM total cross section as a function of the
collision energy (excitation function) for the S($^1$D) + H$_2$($v=0, j=0$) reaction calculated on the
RKHS PES (red line) and DMBE/CBS PES (blue line). The upper energy scale is in K (1 meV$\sim$ 11.6 K)}
\label{fig1}
\end{center}
\end{figure}

Fig.~\ref{fig1} depicts the accurate QM results for the 
excitation function on the two PESs. Although both curves 
have a similar shape and decay rapidly with the collision 
energy, the value of the cross section given by the RKHS 
PES is about three times larger than that obtained on the 
DMBE/CBS. Remarkably, while the excitation function 
corresponding to RKHS displays a series of wide 
oscillations, as noted in our previous work on the system 
\cite{Lara_2010,Lara2_2011}, the results corresponding to 
the DMBE/CBS PES are in general smoother and only below 2 
meV give rise to sharp peaks.

\begin{figure}
\begin{center}
\includegraphics[width=0.8\columnwidth]{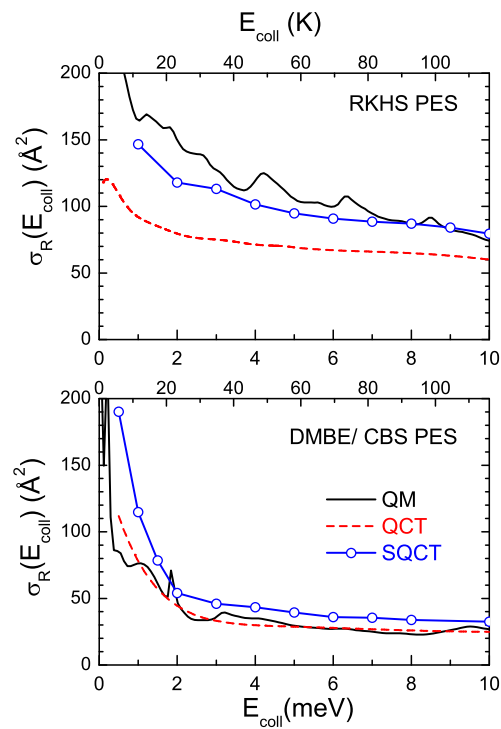}
\caption{(color online) Comparison of the excitation functions calculated using the QM, QCT, and SQCT methods.
Top: Results on the RKHS PES. Bottom: Results on the DMBE/CBS PES. Solid (black) line, QM data;
dashed (red) line, QCT results; solid (blue) line with open circles, SQCT results. The upper energy scale is in K.}
\label{fig2}
\end{center}
\end{figure}

The comparison of the QM results with the excitation 
functions calculated using the other two dynamical 
approaches is shown in Fig.~\ref{fig2}. For the DMBE/CBS 
PES (bottom panel), the QCT cross sections reproduce fairly 
well the smoothed out QM data for energies above 1 meV. As 
expected, the sharp structures observed below 5 meV, likely 
caused by long--lived resonances, are not accounted for by 
the classical calculations. In turn, the SQCT results are 
consistently higher than the QM data in the whole energy 
range. As in previous studies, the difference between QCT 
and SQCT results can be attributed to the failure of  
ergodicity for trajectories that overcome the centrifugal 
barrier but do not form a long--lived complex due to an 
inefficient transfer of momentum to the H$_2$ molecule. As 
can be seen, this difference is relatively small for the 
title reaction, although somewhat smaller on the DMBE/CBS 
PES. For the RKHS PES (upper panel) the SQCT cross sections 
are much closer to the exact QM results than those obtained 
with the QCT calculations. The fact that SQCT calculations 
underestimate the reactivity could suggest a significant 
presence of tunneling, as it will be discussed hereinafter. 
In spite of the differences between the results obtained 
with the QM, QCT and SQCT methods in each PES, all the 
calculations using any of these theoretical approaches 
predict a lower reactivity on the DMBE/CBS PES. Therefore 
it can be concluded that the major dynamical differences 
between the two PES cannot be attributed simply to quantum 
effects.

Further information on the remarkably different reactivity 
on the two PESs can be gained by inspection of the opacity 
functions; {\em i.e.}, the reaction probability as a 
function of the total (here also orbital) angular momentum, 
$P(J)$. Fig.~\ref{fig3} portrays the opacity functions 
obtained using the three methods for 5 meV, 10 meV and 30 
meV collision energies. The differences in shape and 
magnitude between the opacity functions calculated on both 
PES are conspicuous. Not only the highest values of $J$ 
accessible, $J_{\max}$, on the RKHS PES are considerably 
larger in the energy range here considered, but also the 
quantum reaction probabilities in the DMBE/CBS PES are 
smaller  what justifies the big difference in the 
excitation functions. Similar effect occurs with the QCT 
and SQCT calculations except for the lower $J$ values. 

It is convenient to describe the results on the two PESs 
separately. Starting with the RKHS PES, the QM opacity 
function exhibits two regions depending on the $J$ value. 
In the first one, the reaction probability oscillates 
around a high value, $\sim$0.9, suggesting a barrierless 
reaction, in which the energy is much bigger than the 
centrifugal barrier. In the second region, the opacity 
function decreases, manifesting a progressive hindrance for 
the reaction, until the reaction probability vanishes due 
to the inability to overcome the centrifugal barrier. The 
behavior of the SQCT opacity function is somewhat more 
complicated. The results corresponding to the higher 
collision energy, 30 meV, show also two regions but with 
different origin and located at different values of $J$ 
than in the QM case. For low angular momenta, the opacity 
function has a value which is close to unity, but then, 
around $J$=9, gives rise to a sort of step at a constant 
value $\sim$0.7. Finally, by $J$=20, it briskly drops, to 
die at the same $J_{\max}$ as the quantum opacity function. 
However, at $E_{\rm coll}$=5 meV and 10 meV, the first 
region with $P(J)\approx$1 is absent in the SQCT opacity 
functions which takes a constant value of 0.7 until it 
suddenly drops. Interestingly, the values of the SQCT 
probabilities in this region seem to be a lower limit for 
the oscillations of the QM ones. In contrast, in the range 
of $J$ values of the second region where the QM probability 
decreases, the SQCT and QCT probabilities are bigger than 
the QM ones. As for the QCT results, the opacity functions 
exhibit essentially the same shape as the SQCT ones but 
with somewhat smaller values, analogously to what happened 
for the excitation functions. In particular, one can 
distinguish two regions at 30 meV but just one for lowest 
energies, as in the SQCT case. 

The opacity functions on the DMBE/CBS PES are in strong 
contrast with those found on the RKHS PES. In general, they 
are smoother, decreasing gradually with increasing $J$ 
values, after an initial slight increase of the reaction 
probabilities. It is surprising that even for the lowest 
partial waves, the QM reaction probability at 5 meV is 
limited to $\approx$0.5 whereas for those calculated on the 
RHKS PES it was close to unity. In addition, for all the 
three energies, the values of $J_{\max}$ are always lower 
than those corresponding to the calculations on the RKHS 
PES. 

Summarizing, the main dynamical differences between the two
PES in the considered energy range are: (a) The reactivity
corresponding to the DMBE/CBS PES is much lower than that
corresponding to the RKHS PES. In spite of the presumed
barrierless character of the former, it seems that there is
something that systematically hinders the reaction, even
for low angular momenta where the centrifugal barrier is
absent or small. (b) The shapes of the corresponding
opacity functions are strikingly different on the two PESs.
(c) The QM opacity functions and cross sections on the
DMBE/CBS PES seem to oscillate around their QCT
counterparts, and are in better agreement with each other
than in the case of the results on the RKHS PES.

\section{Discussion}

As shown in a previous study~\cite{Lara2_2011}, the 
reaction probability in the title reaction is essentially 
limited by the capture probability in the S+H$_2$ 
arrangement channel. In the energy range here considered, 
there is only one single open channel ($v$=0, $j$=0, 
$\Omega$=0) in the S+H$_2$ arrangement {\em vs.} 16 open 
rovibrational levels in the products with their 
corresponding $\Omega$ states. Consequently, according to 
the statistical model, which assumes the formation of an 
intermediate complex, the probability of complex breakdown 
into products is much larger than into reagents, and 
therefore the reaction probability is essentially equal to 
the capture probability from the initial state. We shall 
prove that indeed the reactivity is dominated by the 
dynamical features of the entrance channel and provide a 
neat explanation of the differences observed in the 
calculations on the two PES. 

Let us start discussing the QM results on the RKHS PES and 
their comparison with the SQCT data. As it has been shown, 
the QM opacity function has a marked oscillating behavior 
for all the energies here considered. These oscillations 
were also found when examining the reaction probabilities 
as a function of the collision energy (not shown here) for 
a fixed $J$ \cite{Banares02}. Even for $J=0$, where there 
is no centrifugal barrier, reaction probabilities in this 
system do not reach the unity in the considered energy 
range except for very particular energies, but quickly 
oscillate around an average of $\approx$0.9. This behavior 
is a manifestation of an indirect mechanism mediated by 
resonances which is characteristic of the presence of the 
deep well supporting the SH$_2$ collision complex. Except 
for the fact that the oscillations become somewhat more 
shallow and that, on average, the reaction probability is 
closer to one, the effect of raising $E_{\rm coll}$ does 
not bring any substantial change with respect to the lower 
energies here examined. 

In contrast, as commented on above, the SQCT results at the 
two lowest $E_{\rm coll}$ here presented give rise to a 
very flat reaction probability, essentially constant with 
$J$, up to a value wherein $P(J)$ drops suddenly and 
becomes zero. Interestingly, the reaction probability in 
this flat region is $\approx$0.7. The fact that the SQCT 
approach underestimates the reactivity in this range of $J$ 
as compared to the QM method indicates that some of the 
incoming flux is classically reflected before the capture 
can take place, while in the QM case those collisions lead, 
however, to complex formation. This behavior can only be 
explained by the contribution of tunneling. 

In order to explain such behavior it is pertinent to 
examine the topography of the entrance channel of the two 
PESs. We have already stated that the big differences in 
reactivity when comparing both PESs are a consequence of a 
significant difference in the $J_{\max}$ value of the 
opacity functions in both QM and classical results, 
together with lower reaction probabilities in the case of 
the DMBE/CBS PES in the whole range of $J$. However, the 
difference in the reactivity cannot be associated with a 
dynamical behavior inside the potential well as long as the 
SQCT and QCT calculations render similar opacity functions 
and, in particular, identical values of $J_{\max}$. Since 
trajectories in the SQCT calculations do not explore the 
well, the discrepancy between the results on the two PESs 
must be primarily due to features located in the entrance 
valley before the deep insertion well. 

Fig.~\ref{fig4} displays contour plots of the entrance 
channel of the PESs for specific H--H internuclear 
distances, $r$. The selected $r$ values correspond to the 
H$_2$ equilibrium distance, $r_{\rm eq}$, and to the inner, 
$r_{-}$, and outer, $r_{+}$, classical turning points for 
the ($v$=0,$j$=0) and ($v$=1,$j$=0) rovibrational states. 
Cuts are drawn as a function of $R_x$ and $R_y$, where 
$R_x$ and $R_y$ are the projections of the Jacobi 
atom--diatom distance, $R$, onto the three atom plane, with 
the $x-$ axis along the H$_2$  molecular axis. 
Specifically, the left panels correspond to the RKHS PES 
and the right panels to the DMBE/CBS PES. Common to all 
these plots is the presence of the insertion well (negative 
green contours) at perpendicular geometries centered at $R 
\approx$ 1.4 \AA. Red contours indicate the repulsive, 
inner part of the potential. In addition, the presence of a 
relatively small barrier at the collinear configuration can 
be observed in most of the plots, and it seems especially 
significant at the H$_2$ equilibrium distance. Although 
there are general resemblances between the two PESs at the 
various H$_2$ distances, there are features which are 
clearly different. In the case of the RKHS PES, for $r < 
r_{\rm eq}$, the collinear barrier is clearly broader than 
that for $r = r_{\rm eq}$, whereas for $r > r_{\rm eq}$, 
the width decreases until it finally disappears. In the 
DMBE/CBS PES, the collinear barrier is much more confined 
and persists for all values of $r>r_{\rm eq}$, and, in 
contrast to the RKHS PES, it grows with the H--H bond 
stretching. Most interestingly though, is the fact that for 
configurations implying a compressed H--H bond, the barrier 
grows considerably and covers a broad range of $R$ values 
and orientations of the H$_2$ molecule. Finally, it is 
worth mentioning the existence of a shallow van der Waals 
well (43 meV depth) at $R \approx$3.5 \AA~(light green 
external regions) in the RKHS PES which is not evident in 
the DMBE/CBS PES. As will be seen, both, the barrier and 
the van der Waals well ref.~\cite{Aoiztomas}, play 
important roles in the dynamics at low collision energies.

\begin{figure}
\begin{center}
\includegraphics[width=1.0\columnwidth]{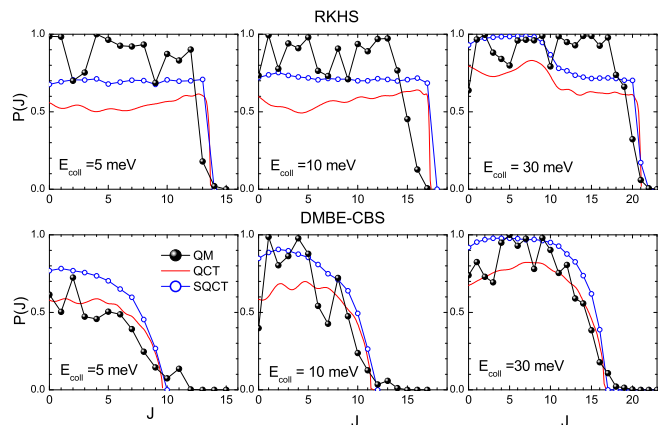}
\vspace{-1cm}
\caption{(color online) Comparison of opacity functions for the S($^1$D)+
H$_2$($v$=0, $j$=0)$\to$SH + H reaction calculated using
accurate QM (black solid lines with filled circles), QCT (red line)
and SQCT (blue line with open circles) approaches at three
specified kinetic energies. Upper panels correspond to the
results obtained using the RKHS PES. Lower ones correspond
to the DMBE/CBS PES.}
\label{fig3}
\end{center}
\end{figure}

\begin{figure}
\begin{center}
\includegraphics[width=0.7\columnwidth]{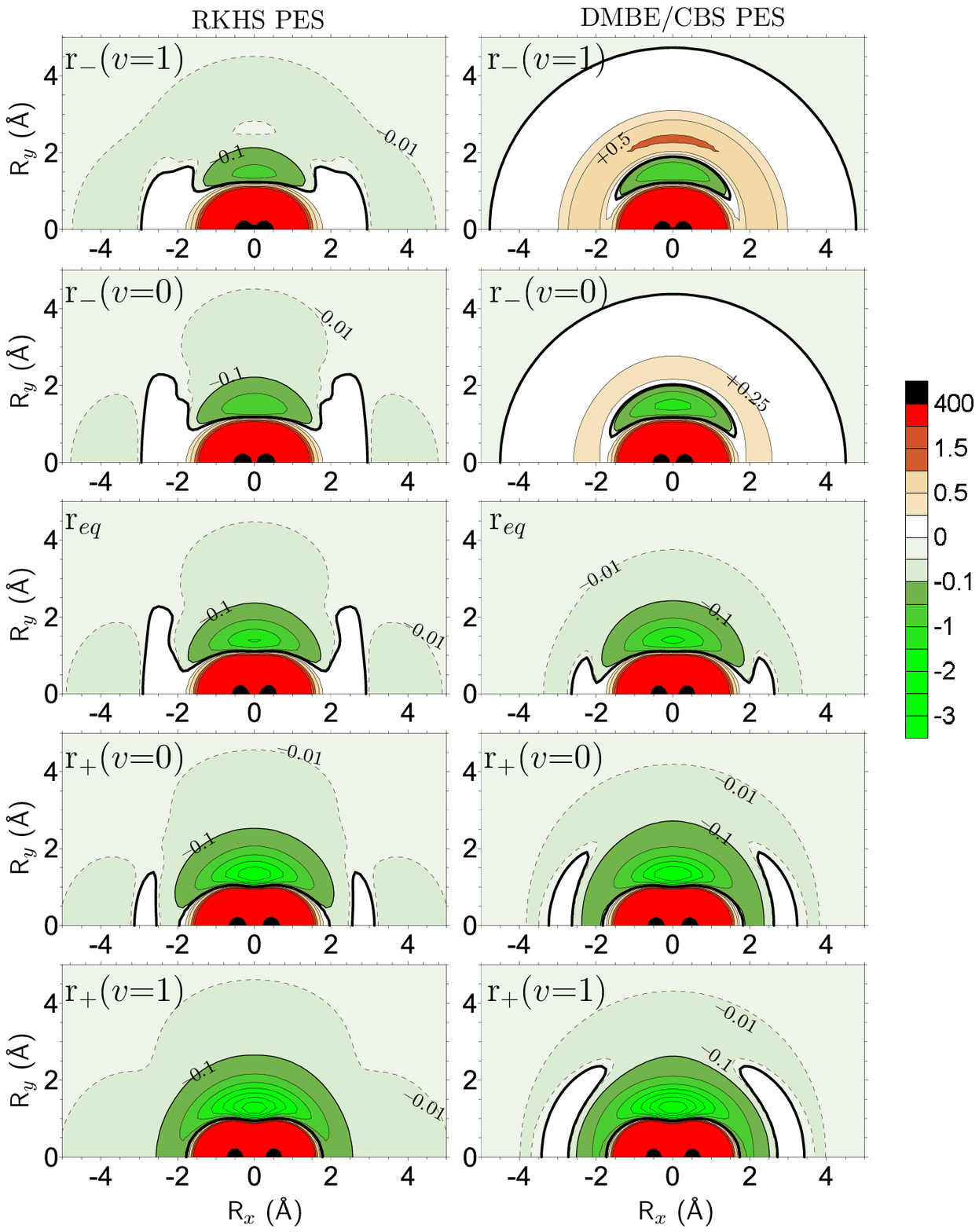}
\caption{(color online) Contour plots of the RKHS PES (left column) and DMBE/CBS PES (right column)
at the H--H internuclear distances corresponding to the inner turning points, $r_{-}(v=1)$ (top row),
$r_{-}(v=0)$ (second row), equilibrium distance, $r_{eq}$, (third row), outer turning points, $r_{+}(v=0)$ 
(fourth row), and $r_{+}(v=1)$ (bottom row). $R_x$ is the components of the $\bm R$ Jacobi vector
($R_{{\rm S}-{\rm H}_2}$) along the internuclear $x$ axis. For each plot the zero of the energy scale 
corresponds to the asymptotic reactant's valley at the indicated H--H distance (0 for $r_{eq}$, 0.27 eV for 
$r_{\pm}(v=0)$ and 0.78 eV for $r_{\pm}(v=1)$). Green contours indicate energies below the 
minimum of the S--H$_2$ valley. Red contours represent highly repulsive parts of the PES. The thick black contour 
represents the zero energy value that limits the ``ear--like'' barrier at collinear configurations. 
Notice the nearly isotropic barrier that appears in the DMBE/CBS PES for the compressed H--H bond.} 
\label{fig4} 
\end{center}
\end{figure}

\begin{figure}
\begin{center}
\includegraphics[width=1.1\columnwidth]{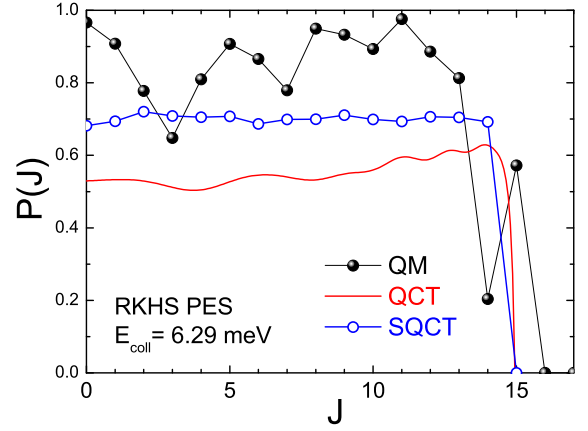}
\caption{(color online) Comparison of QM, QCT, and SQCT opacity functions calculated on the RKHS PES at 
$E_{\rm coll}$=6.26 meV (=73 K). At this energy a shape resonance can be observed for $J$=15. 
See Fig.~\ref{fig1} and ref.~\onlinecite{Lara2_2011}. Lines and symbols as in Fig.~\ref{fig3}.}
\label{fig5}
\end{center}
\end{figure}

Returning to the discussion of the dynamics on the RHKS 
PES, tunneling through the aforementioned barrier is likely 
to occur. For high angular momenta, with significant 
centrifugal barriers, tunneling through the combined 
dynamical and centrifugal barrier gives rise to shape 
resonances which survive the sum over partial waves and can 
be observed at some specific energies, causing marked 
oscillations in the excitation function 
\cite{Lara_2010,Lara2_2011}. A nice example of these shape 
resonances is depicted in Fig.~\ref{fig5} at $E_{\rm 
coll}$=6.26 meV, at which a marked oscillation in the RKHS 
excitation function takes place (see Fig. \ref{fig1}). The 
raise of the $P(J)$ for $J$=15, which is classically 
closed, leads to a substantial increase in the cross 
section. Therefore, although the values of $J_{\max}$ for 
QM results has been shown to usually coincide with those  
found in the SQCT or QCT calculations (see 
Fig.~\ref{fig3}), this is not the case at some specific 
energies, as that shown in Fig~\ref{fig5}. This is a clear 
case of a genuine quantum effect that none of the classical 
approaches can account for. 

An interesting finding is the step for $J$=9 in the SQCT 
$P(J)$ at $E_{\rm coll}$=30 meV, which marks the transition 
from $P(J)$=0.7 to $P(J)$=1.0 (see Fig.~\ref{fig3}). This 
step is absent at lower collision energies wherein the 
maximum value of $P(J)$ is always 0.7. The observation of 
the plots of Fig.~\ref{fig4} provides an explanation for 
this behavior. As can be seen, at the low collision 
energies considered, the approach to the deep attractive 
well is only impeded by the small ``ear-shape'' barrier 
near collinearity at $R\approx$2.5 \AA. Its net effect is 
to partially screen the well making the reaction 
probability be less than one. That is, at sufficiently low 
$E_{\rm coll}$ the cone of acceptance is limited to those 
configurations for which the dynamical barrier is null, 
what approximately amounts 70\% of the collisions. This 
explains why $P(J)$ is flat for a given range of $J$, until 
a value for which the centrifugal barrier is higher than 
the collision energy, and then it drops to zero. The 
appearance of a region where $P(J)$=1 with increasing 
energy (at $E_{\rm coll} \geq 30$ meV) clearly indicates 
that the radial energy is sufficient to overcome the 
dynamical barrier and the cone of acceptance practically 
covers 100\%. Nevertheless, with increasing $J$, and thus 
with a higher centrifugal barrier, the situation reverts to 
that observed at lower energies; namely, there are 
directions of approach for which the reaction cannot take 
place thus reducing the reaction probability. 

The above discussion serves to explain the classical 
behavior. However, the QM reaction probabilities cannot be 
accounted for by the same rationale. Apart from the 
oscillations and the energetically localized resonances as 
that shown in Fig.~\ref{fig5}, the $P(J)$ have consistently 
higher values (that can reach the unit for some $J$ values) 
than those calculated by the SQCT method, and, in fact, the 
SQCT provides a lower limit for the QM reaction 
probabilities. In addition, somehow surprisingly, the fall 
of the QM $P(J)$ is more gradual than in the classical case 
and leads to the apparent paradox that for high $J$ the 
SQCT (and QCT) $P(J)$'s are higher than the QM ones in a 
sort of paradoxical ``anti--tunneling'' behavior. 

\begin{figure}
\begin{center}
\includegraphics[width=0.8\columnwidth]{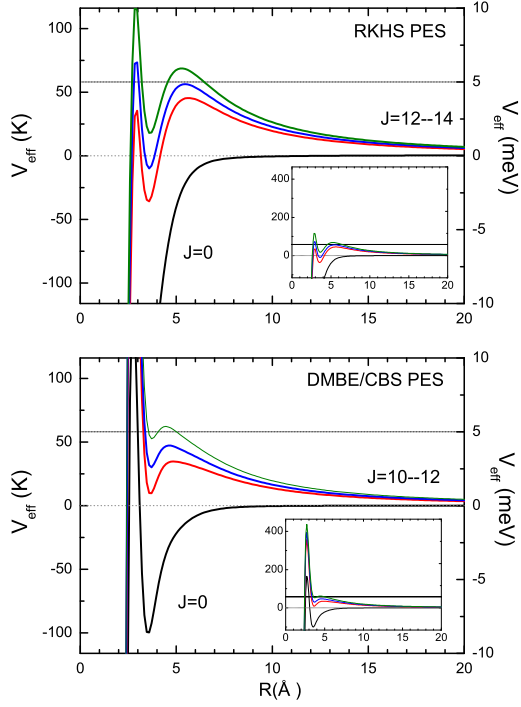}
\caption{(color online) Effective potentials, $\langle  V \rangle_{l,l}(R)+l(l+1) \hbar^2 /2 \mu R^2 $, averaged over
the internuclear distance and orientation of H$_2$($v$=0,$j$=0) molecule.
Upper panel: results on the RKHS PES. Lower panel: results on the DMBE/CBS PES.
Note the presence of a sharp inner barrier and another outer barrier at larger distances.
The minimum between these barriers in RKHS PES may support shape resonances.
In the case of surface DMBE/CBS, the inner barrier is very high and it
is present even for $J=0$ (see inset).}
\label{fig6}
\end{center}
\end{figure}

The situation becomes clear in the light of the 1D quantum 
effective potentials, $\langle  V(R) \rangle_{l,l}+l(l+1) 
\hbar^2 /2 \mu R^2 $, calculated as indicated in 
Section~\ref{QMmethod}. It must be stressed that they are 
constructed by averaging the $V^0(R,r,\theta)$ over $r$ and 
$\theta$ using the rovibrational state as a probability 
distribution and adding the centrifugal term. 
Figure~\ref{fig6} portrays the effective potentials for the 
RKHS PES (upper panel) and the DMBE/CBS PES (lower panel). 
Note the presence in general of two barriers, one external, 
relatively blunt, and a very sharp inner one. In the case 
of the RKHS PES, the origin of the external barrier is 
clear as long as it is absent in the $J$=0 effective 
potential, and therefore can be attributed to the 
centrifugal barrier. The inner, sharp barrier has its 
origin in the collinear barrier mentioned above. As a 
matter of fact, this feature would not be present if only 
the insertion approach had been considered. The interplay 
of these two barriers explains the two regimes observed in 
the QM opacity functions. Taking the 5 meV case as an 
example, as long as the effective potential stays below the 
collision energy (until  $J$=12), the system can access the 
well without hindrance and reacts. The QM $P(J)$ reaches 
high probabilities (on average, close to one) in what we 
have called the QM first region. In this regime, the 
external barrier is always larger than the internal one, 
and we can think of it as the `bottleneck' which limits the 
flux entering into the well. However, for $J=13$, it is the 
internal barrier the one which starts blocking the access 
to the well at the considered energy, in coincidence with 
the decay of the opacity function. In this regime, the 
reaction occurs via tunneling through the inner (thin) 
barrier. Finally, for $J=14$, both the external and the 
internal centrifugal barriers are higher than the kinetic 
energy, and the opacity function essentially vanishes. In 
summary, the two regimes of QM opacity functions are 
closely related to the maxima in the effective potential. 
Let us also remark that it is precisely this double maximum 
structure comprising a local minimum, (see Fig.~\ref{fig6}) 
the one which supports the above mentioned shape resonances 
for particular partial waves. These resonances are the 
origin of some of the oscillating structures in the QM 
excitation functions, as commented on above and discussed 
in a previous work~\cite{Lara2_2011}.

The presence of the second, inner, barrier in the effective 
potentials of the RKHS PES sheds light on the somewhat 
surprising ``anti--tunneling'' behavior. In the classical 
picture, a trajectory {\em feels} only the local value of 
the potential at the points the trajectory is going 
through. In the quantum picture, the system is sensitive to 
the average (thus `non local') value of the potential. As 
the potential has to be averaged over the whole angular 
range (and over internuclear distances), the distinction 
between the directions of approach with and without barrier 
looses its meaning. Effectively, except for tunneling, this 
situation would be equivalent to that resulting from a QCT 
calculation with an isotropic potential than would be 
smaller than the local value of the barrier. In that case, 
the SQCT $P(J)$ will be zero just at the $J$ value in which 
the QM ones start decreasing. While classically the 
hindrance in reactivity is understood as a fractional 
closure of reactive configurations, the process can be 
described in quantum mechanical terms as a decreasing 
tunneling through an increasing barrier. 

As shown hereinbefore, the opacity functions calculated on 
the DMBE/CBS (see Fig.~\ref{fig3}) are strikingly 
different, and it seems that the previous arguments could 
not be valid to explain those results. In fact, the SQCT 
reaction probabilities constitute an upper limit for the QM 
results at the three energies examined. Let us consider now 
the effective potential in the case of the DMBE/CBS PES. 
Surprisingly, the presence of a very high and sharp 
internal barrier could limit the access to the well even 
for $J=0$. This dynamical internal barrier, in an 
apparently {\em barrierless} reaction, must be the reason 
for the much lower QM reaction probabilities. Its origin 
will be justified below. Comparing the QM $P(J)$ at 5 meV 
with the plot of the effective potential, it becomes clear 
that it is the height of the external barrier what 
determines the $J_{\max}$ accessible for the reaction to 
take place. For a given $J$, this external barrier is 
always lower for the RKHS PES than for the DMBE/CBS PES, 
what justifies the larger $J_{\max}$ in the former. In 
addition, the inner barrier is much higher in the case of 
the DMBE/CBS PES even for $J$=0, and becomes broader with 
increasing $J$ to the point of collapsing with the outer 
barrier into a single crest at $J \approx$15 (not shown). 
In addition, for $J < 15$ the well between the maxima is 
more shallow than in the RKHS PES. The features of the PES 
responsible of the inner barrier have to be the origin of 
the also much smaller reactivity in the classical 
treatments. 

At this point, we need to return to Fig.~\ref{fig4} to 
relate the topography of the PES to the two remarked 
features of the effective potentials: (i) the more 
pronounced minimum between maxima in the case of the RKHS 
PES, responsible for the higher $J_{\max}$, and (ii) the 
high inner barrier present in the DMBE/CBS PES, responsible 
for the much smaller values of reaction probabilities even 
for $J=0$. 

With regard to the feature (i), it can be shown that the 
minimum in the effective potential of the RKHS PES relies 
on the aforementioned van der Waals, mostly absent in the 
DMBE/CBS PES. Far from being an artifact of the fit, this 
well seems to be real, appearing also in some {\em ab 
initio} calculations performed to verify its 
existence~\cite{Lara2_2011}, although with smaller depth. 
It is also the origin of many of the structures in the 
cross-sections of the RKHS PES and of the decrease of 
height of the centrifugal barriers. 

As for the feature (ii), the origin of the high inner 
barrier in the effective potential of the DMBE/CBS PES can 
be traced back to the nearly isotropic barrier that appears 
as the H--H bond starts to be compressed. This barrier 
covers a broad range of $R$ between 1.5 and 4.5 \AA, as 
shown in Fig.~\ref{fig4}, and its height is strongly 
dependent on the value of $r$. For T-shape configuration it 
may become very significant; as high as 200 meV at the 
inner turning point for $v$=0 and surpassing 1 eV for the 
inner turning point of $v$=1. This remarkable feature is 
basically absent in the RKHS PES. 

The much lower reactivity of the DMBE/CBS can thus be 
attributed to this barrier to the extent that actually the 
global PES is not truly barrierless. As a result of this, 
the reaction probability calculated on this PES decreases 
gradually with the impact parameter, as shown in 
Fig.~\ref{fig3}, and the shape of the opacity functions at 
sufficiently low collision energies calculated by any of 
the three methodologies deviates considerably from what can 
be expected for a typical barrierless reaction, as typified 
by the corresponding results on the RKHS PES. It must be 
stressed that the mentioned barrier only appears for 
compressed H--H bond configurations and it maximum height 
turns up at $R \approx$2.2 \AA.  

\begin{figure}
\begin{center}
\includegraphics[width=0.8\columnwidth]{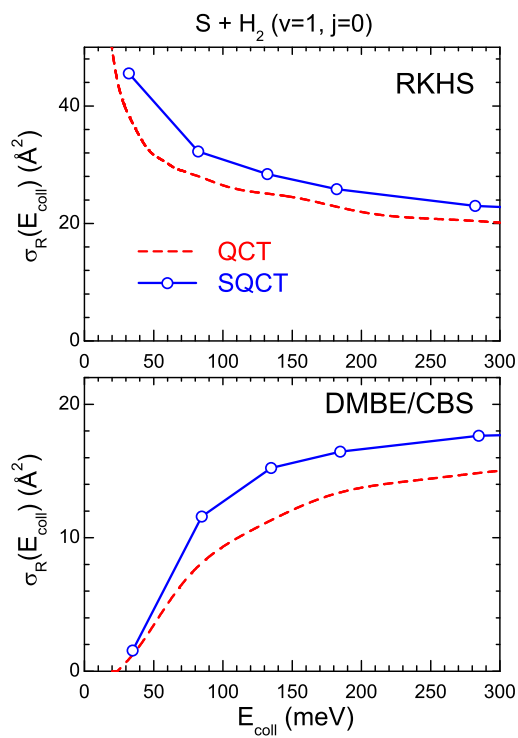}
\caption{(color online) Comparison of excitation functions for the S($^1$D)+
H$_2$($v$=1, $j$=0)$\to$SH + H reaction calculated with the QCT (red, dashed line) and SQCT
(blue open circles) approaches on the RKHS PES (top) and DMBE/CBS PES (bottom). 
Note the different scales in each plot.}
\label{fig7}
\end{center}
\end{figure}

It can be expected that the effect of this barrier will 
show up blatantly for the reaction with vibrationally 
excited molecules for which the H$_2$ internuclear distance 
can reach shorter values. In particular, the corresponding 
QCT and SQCT excitation functions for the reaction with 
H$_2$($v$=1, $j$=0) are shown in Fig.~\ref{fig7} in the 
25--300 meV collision energy range. The $\sigma_R(E_{\rm 
coll})$ calculated on the RKHS PES follows the expected 
behavior: a shape, monotonically decreasing with $E_{\rm 
coll}$, and absolute values similar to those obtained in 
the reaction with with H$_2$($v$=0, $j$=0). In strong 
contrast but at this point not surprisingly, the QCT and 
SQCT excitation functions calculated on the DMBE/CBS 
exhibit a threshold at $E_{\rm coll} \approx$ 25 mV. The 
shape of the excitation functions is akin to that of a 
reaction with barrier. Only at sufficiently high $E_{\rm 
coll}$ the values of $\sigma_R(E_{\rm coll})$ become 
comparable to the results found for the reaction with 
H$_2$($v$=0, $j$=0). 

\begin{figure}
\begin{center}
\includegraphics[width=1.0\columnwidth]{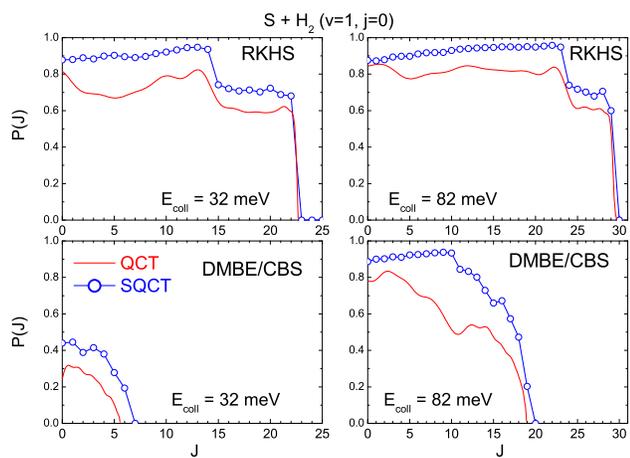}
\caption{(color online) Opacity functions for the reaction with H$_2$($v$=1,$j$=0) at 32 meV (left panels)
and 82 meV (right panels) collision energy calculated with the QCT and SQCT
approaches on the RKHS PES (top panels) and DMBE/CBS PES (bottom panel).
These collision energies correspond to the first two points of the excitation functions
of Fig.~\ref{fig7} calculated with the SQCT method. Lines and symbols as in Fig.~\ref{fig3}.}
\label{fig8}
\end{center}
\end{figure}

The opacity functions corresponding to 32 meV and 82 meV 
calculated on both PESs are shown in Fig.~\ref{fig8}. The 
results obtained on the RKHS (upper panels) are similar to 
those found at $E_{\rm coll}$=30 meV for $v$=0, $j$=0, 
displaying two distinct regimes; a first one with higher 
reaction probabilities $\approx 0.9$ and, a second one that 
appears as a notch before the sudden drop to zero. It is 
worth noticing that the maximum value of the SQCT $P(J)$ is 
somewhat smaller than that obtained in the reaction with 
H$_2$ in $v$=0 due to a larger probability of complex 
breakdown into reactants as the total energy increases. The 
results on the DMBE/CBS PES clearly indicate lower reaction 
probabilities than in the case of the $v$=0. Interestingly, 
vibrational excitation does not promote but hampers the 
reactivity. 

Let us finally note that as stated in 
ref.~\onlinecite{Var}, the grid of {\em ab initio} points 
used for the construction the DMBE/CBS PES sampled the 
S-H$_2$ entrance channel for H--H internuclear distances in 
the $1.4 {\rm a}_0 \le r \le 3.4 {\rm a}_0$ range. In view  
of this sampling, and being the equilibrium distance of 
H$_2$ precisely $1.4$ a$_0$ we may conclude that the 
observed repulsive barrier is likely an artifact of the fit 
of the PES, perhaps reflecting the lack of electronic 
structure calculations for short H--H internuclear 
distances. This work is then illustrative of the 
conflicting objectives that are present when modeling a 
PES. If the electronic structure calculations can be done 
accurately and expediently, numerical or semi-numerical 
interpolation schemes can be utilized to reproduce the 
subtle details that govern nuclear dynamics on 
single-sheeted PESs, as is the case with the RKHS $\rm 
H_2S$ PES. This is so, even if extrapolation to the 
asymptotes can somewhat endanger the analysis. However, 
when accurate calculations are just hardly affordable, then 
a least-squares fit based whenever possible on a physically 
motivated form like DMBE that can be calibrated from a 
relatively small number of {\em ab initio} points may be 
ideal. Such a dilemma poses itself a compromise that has 
been well illustrated in the present case study. In fact, 
the major goal of Ref.~\onlinecite{Var} has been to test, 
to our knowledge for the fist time, a general 
cost-effective methodology~\cite{VAR07:244105,VAR07:398} to 
generate accurate global PESs using traditional correlated 
{\em ab initio} methods and basis sets with low cardinal 
numbers. The focus has then been on the quality of the 
calculated energies, by limiting their number to a minimum 
level, while relying mostly on the predicting capability of 
the DMBE formalism to generate the global PES. Clearly, if 
subtle topographical features are at stake, the method 
should be used with caution. This can be done either via a 
gradually increase of the size of the grid of {\em ab 
initio} points until definite trends are observed or by 
joint use of dynamics calculations, which may hopefully 
suggest additional {\em ab initio} calculations by alerting 
for critical predictions and hence for the sparseness of 
the grid at specific regions of configuration space 
\cite{Collins:jcp95,Collins:jcp98}. It should be recalled 
that, interestingly, fits based on similar forms but using 
slightly different grid sizes may lead to different 
results, mostly when possessing excessive flexibility. In 
fact, even if such differences can be minor and hardly 
visible, they can be made to play a key role if specific 
energy regimes are devised, such as the case with the low 
collision energies here considered. Indeed, this has been 
exemplified for $\rm H_2S$, with the fitting artifacts 
reported here (and elsewhere~\cite{MAR010,MAR011}) for the 
DMBE/CBS PES. These features essentially disappear when 
considering the DMBE/SEC PES~\cite{Var2}, in spite of the 
fact that both basically employ the same grid of {\em ab 
initio} energies as calibration data. Thus, the release of 
a global PES prior to use with a sufficiently demanding 
test set of dynamics calculations runs the risk of failure 
whenever a property outside the test set is considered. 
This has been the case with the DMBE/CBS PES for which a 
recalibration is desirable. Similarly, even if they are 
seemingly small, a careful evaluation of non-adiabatic 
effects for the title system is imperative thus requiring 
the modeling of a multi-sheeted PES.

\section{Conclusions}

Detailed quantum mechanical (QM), quasi--classical (QCT) 
and statistical (SQCT) calculations have been performed for 
the S($^1D$)+H$_2$($v$=0, $j$=0) insertion reaction on two 
different PESs in the 0.09--10 meV collision energy range. 
Although the differences between the results obtained with 
the three methodologies are appreciable, the QCT and SQCT 
are in a general good agreement with the accurate single 
PES QM calculations. In particular, the good agreement 
between SQCT and the accurate QM calculations lends 
credence to the tenet that the reaction is limited by the 
capture probability in the entrance channel, and that the 
formation of a the collision complex takes place once the 
centrifugal barrier is surmounted. As such, for this type 
of reactions, the study of collisions at very low 
translational energies may be a sensitive probe of the 
detailed topography of the entrance channel and the 
long--range part of the PES.

Even though it can be expected that the amount of reactive 
flux should be mostly determined by the purely centrifugal 
barrier, rather subtle topographical features localized in 
specific regions of the PES may influence on the reactivity 
at low energies. Indeed, the three sets of calculations 
show unequivocally that the reactivity on the RKHS PES is 
much larger than that obtained on the more recent DMBE/CBS 
PES. In fact, the opacity functions determined on the two 
PESs are remarkably different in shape and in absolute 
value, and reveal that the lower reactivity on the DMBE/CBS 
PES is due to lower values of the reaction probability in 
practically the whole range of orbital angular momenta (or 
impact parameters), as well as to considerably smaller 
values of maximum impact parameter leading to reaction. 

The detailed inspection of the opacity functions and the 
comparison of the SQCT and QM results provides important 
clues on the detailed dynamics of the reaction on each PES. 
As a matter of fact, the shape and absolute values of the 
reaction probabilities have been explained in terms of the 
topography of the PES and, in particular, of the 
corresponding effective potential in the entrance channel, 
which are averaged over the internuclear distances and 
orientations of the H--H molecule. In the case of the RKHS, 
it has been shown that the small collinear barrier, whose 
influence at higher energies is very minor, plays a 
significant role on the features found in the opacity 
functions at small collision energies. In the case of the 
the DMBE/CBS PES, the huge differences found in the 
reaction probabilities with respect to those on the RKHS 
PES can be attributed to the presence of a barrier that 
becomes very significant when the H--H bond is compressed, 
a region of the PES wherein {\em ab initio} data were 
largely missing. 

To confirm the hampering role for the reaction to occur 
associated with this barrier at small H--H internuclear 
distances, QCT and SQCT calculations of cross sections and 
opacity functions were also carried out for vibrationally 
excited H$_2$ in $v$=1. Whilst on the RKHS the excitation 
function is analogous to that found for $v$=0, in the case 
of the DMBE/CBS not only the cross-section for $v$=1 is 
lower than that for $v$=0 but also gives rise to a reaction 
threshold, a clearly unexpected feature for a barrierless 
reaction. 

This work shows that the study of presumably barrierless 
reactions at low collision energies ($<100$K) can reveal 
features in the global PES that would have been unnoticed 
at higher energies. These features can and indeed do affect 
the dynamics of the reaction whose accurate description is 
mandatory to undertake studies at the cold and ultracold 
regimes as it has long been anticipated \cite{VAR88:59} but 
not demonstrated as clearly as here done via rigorous 
classical and quantum dynamics calculations on accurate 
{\em ab initio} based PESs. Indeed, the present study can 
be illuminating about the difficulties encountered in 
modeling an all-purpose global potential energy surface.

\vspace{1cm}  
\noindent {\bf Acknowledgments}\vspace{2ex}

The support of the Spanish Ministry of Science and 
Innovation (grants CTQ2008-02578/BQU and Consolider Ingenio 
2010 CSD2009-00038) are gratefully acknowledged. PGJ 
acknowledges the FPU fellowship AP2006-03740. AJCV thanks 
the support of Fundação para a Ciência e a Tecnologia, 
Portugal.

\newpage

\newpage

\newpage

\newpage

\newpage

\newpage

\clearpage

\end{document}